\documentclass[10pt,aps,pre,twocolumn]{revtex4}
\usepackage[active]{srcltx}
\usepackage[utf8]{inputenc}
\usepackage{latexsym}
\usepackage{amsmath,amssymb,bbm,braket,bm}
\usepackage{graphicx}
\usepackage[hypertex]{hyperref}
\usepackage{ulem}
\usepackage{color}

\def\:={\,\raisebox{0.85pt}{.}\hspace{-2.78pt}\raisebox{2.85pt}{.}\!\!=\,}
\def\=:{\,=\!\!\raisebox{0.85pt}{.}\hspace{-2.78pt}\raisebox{2.85pt}{.}\,}

\begin{document}

\title{Modulated Phases in a 3D Maier-Saupe Model with Competing Interactions}

\author{P. F. Bienzobaz}
\email{paulabienzobaz@uel.br}
\affiliation{Departamento de F\'isica, Universidade Estadual de Londrina, Caixa Postal 10011, 86057-970, Londrina, PR, Brasil }
\affiliation{Department of Physics, Boston University, 590 Commonwealth Avenue, Boston, Massachusetts 02215, USA}%

\author{Na Xu}
\email{naxu@bu.edu}
\affiliation{Department of Physics, Boston University, 590 Commonwealth Avenue, Boston, Massachusetts 02215, USA}

\author{Anders W. Sandvik}
\email{sandvik@buphy.bu.edu}
\affiliation{Department of Physics, Boston University, 590 Commonwealth Avenue, Boston, Massachusetts 02215, USA}
\date{\today}

\begin{abstract}

This work is dedicated to the study of the discrete version of the Maier-Saupe model in the presence of competing interactions. 
The competition between interactions favoring different orientational ordering produces a rich phase diagram
including modulated phases. Using a mean-field approach and Monte Carlo simulations, 
we show that the proposed model exhibits isotropic and nematic phases, and also a series of modulated phases that 
meet at a multicritical point; a Lifshitz point. Though the Monte Carlo and mean-field phase diagrams show some 
quantitative disagreements, the Monte Carlo simulations corroborate the general behavior found within the mean-field 
approximation.

\end{abstract}
\maketitle

\section{Introduction}

\noindent
Many condensed-matter systems, such as magnetic compounds, polymers and liquid 
crystals, exhibit interesting phases with periodic structures \cite{Chaikin}. Microscopically this 
modular behavior can be understood as resulting from competing interactions favoring different 
orderings \cite{Selke}. Perhaps the most simple and interesting example is the ANNNI 
(axial-next- nearest-neighbor-Ising) model, with competing interactions between first (ferromagnetic) 
and second (antiferromagnetic) neighbors along one specific direction  \cite{PerBak}. The phase diagram, as a 
function of the temperature and the parameter regulating the degree of competition between different 
interactions, exhibits a paramagnetic and a ferromagnetic phase, as well as an infinite series of modulated 
phases \cite{Fisher}. All these phases meet at a special critical point called the Lifshitz point  \cite{Hornreich}. 
Because of its rich phase diagram, the ANNNI model has been widely studied using different analytical and numerical 
methods, and it also has experimental applications  \cite{Fisher79,Fisher81,WSelke}.

In the field of liquid crystals (LCs), there is also significant interest in modulated 
phases \cite{Chen1,Chen2, Mohammad}.The constituent molecules of LCs have a rigid part, which 
is responsible for the alignment of the molecules along a direction 
(described by a director, an angle in the range $\left[0, \pi\right]$), and a more flexible 
part which induces the fluidity. The different phases 
in this state of matter depend on the preferential ordering of the molecules, which in 
turn depends on the temperature. Their characterization are given by the underlying translational 
and rotational symmetries, and are usually classified as nematic, smectic, and cholesteric (also known as 
the chiral or helical phase). The uniaxial nematic phases are well established in the phase diagram 
of a large number of LCs \cite{deGennes, Onsager,Maier1,Maier2,Flory,Zannoni,Danilo} as well as the biaxial nematic phases and its stability \cite{Zannoni1,Eduardo,Henriques}. 

In general, statistical formulations defined on lattices describe satisfactorily many physical characteristics of
thermotropic LCs, and their mean-field formulations can describe nematic phases and the related phase transitions
\cite{Onsager,Maier1,Maier2,Flory}. The simplest and most important model is the Maier-Saupe model, which has been
successful not only in explaining the orientational properties, but also hosts an order-disorder transition, i.e., a 
transition between nematic and isotropic phases \cite{Zannoni}. 

In this work, we consider a generalization of the Maier-Saupe model on a 3D lattice which includes competing interactions along one specific 
direction, similar to the ANNNI model. We show that this frustrated Maier-Saupe model hosts a series of modulated phases that may be related to the
cholesteric phase observed in some in LCs \cite{Lubensky, Kiselev}. Here, as a first step to start exploring frustration effects in LCs, instead 
of describing the molecules by continuously-varying vector degrees of freedom, we consider a discrete version first. Similar to the standard
discrete version of the Maier-Saupe model, the 'molecules' are discrete spins that can only take three different orientations. This simiplification 
makes the model equivalent to the 3D 3-state ANNN-potts model. As far as we are aware, even though there has been numerous works reported
on the standard 3D 3-state potts model with nearest-neighbor interactions \cite{Wu,Selke_Wu}, as well as a 2D ANNN-potts model \cite{nijs85}, no 
research has been done on this type of system in 3D. Thus from either the view of understanding the rich phases in LCs or the more general
perspective of enriching our knowledge of potts models, such a study is desirable. In fact, even though this discrete version of Maier-Saupe 
model with competitions can not fully describe the complexity of the cholesteric phase, our results still show that this model produces modulated 
structures, with periodicity that depends on the parameter regulating the competing interactions and the temperature.

The paper is organized as follows: In Sec.~\ref{sec:problem} we introduce the Maier-Saupe model with competing 
interactions. Sec.~\ref{sec:analytic} is dedicated to an analytical calculation of the order parameter and the free energy 
by means of a variational mean-field approach, which is solved numerically to obtain the phase diagram. Monte Carlo (MC) simulations aimed at an unbiased determination of the phase boundaries are discussed in Sec.~\ref{sec:mc},
and final remarks are given in Sec.~\ref{sec:summary}.

\section{Statement of the problem}
\label{sec:problem}

\noindent
To describe a LC statistically, it is appropriate to define an order parameter in terms of a unitary director ${\bf n}$ that 
corresponds the preferential ordering of the molecules. Due the quadrupole symmetry, the LCs are indistinguishable under
${\bf n} \rightarrow - {\bf n}$ transformation and a natural order parameter that takes it into account is given by 
the second-order tensor,
\begin{eqnarray}
T^{\mu\nu}= a~n^{\mu}n^{\nu}+b~\delta_{\mu\nu},
\label{generic_tensor}
\end{eqnarray}
where $a$ and $b$ are arbitrary constants and $n^{\mu}$ are components of the
director ${\bf n}$, with $\mu,\nu=\left\{x,y,z\right\}$. 
The trace of the tensor $T$ does not contain any orientational information, 
and a convenient order parameter to describe a nematic LC, where the molecules 
have axial symmetry, can be defined by eliminating the trace part as
\begin{eqnarray}
S^{\mu\nu}\equiv T^{\mu\nu}-\frac{\delta_{\mu\nu}}{3}~\text{Tr}\{T\}
=\frac{a}{3}\left(3 n^{\mu} n^{\nu}-\delta_{\mu\nu}\right).
\label{order}
\end{eqnarray}
In a lattice model we can use the expectation value of the elements of such a tensor order parameter at an arbitrary site $i$,
\begin{eqnarray}
M^{\mu\nu}=\left<S_{i}^{\mu\nu}\right>=\left<\frac{1}{2}(3n_{i}^{\mu}n_{i}^{\nu}-\delta_{\mu\nu})\right>,
\label{msc2}
\end{eqnarray}
where $n_i^{\mu}$  are the components of the vector that defines the preferred orientation of the molecule located on site $i$. 
In the nematic phase the order parameter must be nonzero and in the isotropic phase it should vanish. 
In this way we choose $a = 3/2$  such that  $M^{\mu\nu}=1$ in the perfectly ordered phase.

The weak first-order transition between uniaxial nematic and
anisotropic phases is well understood from theoretical as well as experimental investigations \cite{Keyes,Egbert,Peter,Jonathan}.
The transition is well described by the mean-field Maier-Saupe theory, and, following this approach, we will here implement competing interactions
in the Maier-Saupe model (also known as the Lebwohl-Lasher model \cite{LL}). In analogy with the ANNNI model, we consider interactions between
first neighbors along the $x$ and $y$ axes and competing interactions along the $z$-axis, and introduce the following Hamiltonian:
\begin{eqnarray}
\mathcal{H}&=&-J_1\sum_{\mu,\nu}\sum_{x,y,z}\left(S_{xyz}^{\mu\nu}S_{x+1yz}^{\mu\nu}
+S_{xyz}^{\mu\nu}S_{xy+1z}^{\mu\nu}\right.\nonumber\\
&+&\left.S_{xyz}^{\mu\nu}S_{xyz+1}^{\mu\nu}\right)- J_2\sum_{\mu\nu}\sum_{xyz}S_{xyz}^{\mu\nu}S_{xyz+2}^{\mu\nu}.
\label{msc4}
\end{eqnarray}
To achieve the desired competition, the couplings between first neighbors, $J_1$, and second neighbors, $J_2$, should have opposite signs. 
The lattice sites will be labeled by a suffix $xyz$, where $1\le x,y,z \le N$; i.e., the total number of sites is $N^3$. The reason for
introducing the spatially anisotropic coupling is that, once a layered structure has formed, there
is no reason for the effective couplings in the simplified lattice model to be isotropic, and the analogy with the ANNNI model suggests that
the frustration only in the interaction between the layers (our $z$ direction) should be the simplest way to achieve the modulated phases.
The anisotropic interaction still also allows for isotropic (disordered) and nematic phases.

A convenient explicit form for $S_{xyz}^{\mu,\nu}$ is given by
\begin{eqnarray}
S_{xyz}^{\mu\nu}=\frac{1}{2}(3n_{xyz}^{\mu}n_{xyz}^{\nu}-\delta_{\mu\nu}),
\label{msc2b}
\end{eqnarray}
and we will proceed using it to calculate the partition function associated with the Hamiltonian (\ref{msc4}). The partition function is
\begin{eqnarray}
\mathcal{Z}=\sum_{\left\{{\bf n}\right\}}\text{e}^{-\beta \mathcal{H}},
\label{msc5}
\end{eqnarray}
where the sum is over all allowed directions of the vectors ${\bf n}_i$ allowed and $\beta= 1/(k_BT)$, with $T$ the temperature and $k_B$ 
Boltzmann's constant (which we set to $1$ henceforth). A considerable simplification, which we will adopt here, is to consider a 
discrete version of the director as proposed by Zwanzig \cite{Zwanzig}, where the site directors can be oriented only along three 
perpendicular directions,
\begin{eqnarray}
{\bf n}_{i}=\left\{
\begin{array}{l}
(0,0,1),\\
(0,1,0), \\
(1,0,0).
\end{array}
\right.
\label{director}
\end{eqnarray}
This approach works very well in the mean field Maier-Saupe model, where the fluctuations are not so relevant for the main features of the phase
diagram, and, despite the discrete simplification, the usual Maier-Saupe model (without competition) shows qualitatively the physical behavior
of LCs \cite{Figueiredo}. Because of the symmetries, the model is also equivalent to a three-state frustrated Potts model \cite{Wu,Selke_Wu}, 
an extension of the standard ANNNI model. While a generalized $S=1$ ANNNI model (i.e., with three states per lattice site) has been previously 
studied \cite{Muraoka}, the symmetries of this model are different. To our knowledge the Potts version of the ANNNI model has not been studied previously.

When we turn on the competing interactions, the calculation of the partition function is a formidable task even with the discrete approximation.
In the next section we will first employ a variational (mean-field) approach to obtain an approximate analytical expression for the order 
parameter and the free energy and obtain the phase diagram numerically. In Sec.~\ref{sec:mc} we apply MC simulations and extract a phase 
diagram and this is in good general agreement with the mean-field version.


\section{Variational Approach}
\label{sec:analytic}

\noindent
Let $F$ be the free energy of the system. To implement
the Bogoliubov variational method, we need to find the free energy $F_0$
corresponding to a trial Hamiltonian $\mathcal{H}_0$, satisfying the inequality
\begin{eqnarray}
F\leq F_0 +\left<\mathcal{H}-\mathcal{H}_0\right>_0\equiv\Phi.
\label{msc7}
\end{eqnarray}
The notation $\left<\right>_0$ represents an average with respect to the partition
function of the Hamiltonian $\mathcal{H}_0$, that can be parameterized as
\begin{eqnarray}
\mathcal{H}_0=-\sum_{xyz}\sum_{\mu\nu}h_z^{\mu\nu}S_{xyz}^{\mu\nu}.
\label{msc9}
\end{eqnarray}
Here $h_z^{\mu\nu}$ is a symmetric tensor which should be considered as a variational  parameter to 
minimize the free energy $\Phi$ in Eq.~(\ref{msc7}). As shown below, from this approach it is possible to obtain 
self-consistent analytical equations for the order parameter.

We start by calculating the partition function $\mathcal{Z}_0$,
\begin{widetext}
\begin{eqnarray}
\mathcal{Z}_0\!\!
&=&\sum_{\left\{{\bf n}_{xy1}\right\}}\!\!\!\!\exp\!\left[\beta
\sum_{\mu\nu}\left(S_{111}^{\mu\nu}+\cdots
S_{NN1}^{\mu\nu}\right) h_1^{\mu\nu}\right]
\times\cdots\times
\sum_{\left\{{\bf n}_{xyN}\right\}}\!\!\!\!\exp\left[\beta
\sum_{\mu\nu}\left(S_{1,1,N}^{\mu\nu}+\cdots
S_{NNN}^{\mu\nu}\right)
h_N^{\mu\nu}\right]\nonumber\\
&=&\prod_{z=1}^{N}\left\{\sum_{{\bf n}_{xyz}}
\exp\left[\beta\sum_{\mu\nu}S_{xyz}^{\mu\nu}h_z^{\mu\nu}\right]\right\}^{N^2}.
\label{msc10}
\end{eqnarray}
\end{widetext}
From this, taking into account that the director ${\bf n}_{xyz}$ can assume 
six different values in accordance with (\ref{director}), we obtain
\begin{eqnarray}
Z_0=\prod_{z=1}^N\left\{2\exp\left[-\frac{\beta}{2}\sum_{\mu}h_z^{\mu\mu}\right]\sum_{\nu}
\exp\left(\frac{3}{2}\beta
h_z^{\nu\nu}\right)\right\}^{N^2}
\label{msc11}
\end{eqnarray}
and, consequently
\begin{eqnarray}
\!\! \! \! \!F_0\!=\!- \frac{N^2}{\beta}\!\! \sum_{z=1}^N\! \left\{\!\ln \!2\!-\!\frac{\beta}{2}\!\sum_{\nu}\!h_z^{\nu\nu}
\!+\! \ln\!\!\left[\!\sum_{\nu}\! \exp\!\left(\!\frac{3}{2}\beta h_z^{\nu\nu}\!\right)\!\right]\!\right\}\!\!.
\end{eqnarray}
In this case, the problem reduces to the calculation of $\left<\mathcal{H}-\mathcal{H}_0\right>$,
\begin{eqnarray}
\left<\!H\!-\!H_0\right>_0\!\!&=&\!\!-J_1\!\!\sum_{\mu\nu}\!\sum_{xyz}\!
\left(\!\left<S_{xyz}^{\mu\nu}S_{x+1yz}^{\mu\nu}\right>_0\!\!
+\!\!\left<S_{xyz}^{\mu\nu}S_{xy+1z}^{\mu\nu}\right>_0\!\right.\nonumber\\
&+&\left.\left<S_{xyz}^{\mu\nu}S_{xyz+1}^{\mu\nu}\right>_0\right)
-J_2\!\sum_{\mu\nu}\sum_{xyx}\left<S_{xyz}^{\mu\nu}S_{xyz+2}^{\mu\nu}\right>_0
\nonumber\\
&+&\sum_{\mu\nu}\sum_{xyz}h_z^{\mu\nu}\left<S_{xyz}^{\mu\nu}\right>_0\!.
\label{msc12}
\end{eqnarray}
We need to determine the averages of right side of the equation (\ref{msc12}).
It is straightforward to show that $\left<S_{xyz}^{\mu\nu}S_{xyz+1}^{\mu\nu}\right>_0=
\left<S_{xyz}^{\mu\nu}\right>_0\left<S_{xyz+1}^{\mu\nu}\right>_0$ and then it is sufficient to
calculate $\left<S_{xyz}^{\mu\nu}\right>_0$;
\begin{eqnarray}
\left<S_{xyz}^{\mu\nu}\right>_0&=&\frac{\displaystyle\sum_{{\bf n}_{xyz}}S_{xyz}^{\mu\nu}
\exp\left(\beta\displaystyle\sum_{\mu\nu}h_z^{\mu\nu}S_{xyz}^{\mu\nu}\right)
}{\displaystyle\sum_{{\bf n}_{xyz}}
\exp\left(\beta\displaystyle\sum_{\mu\nu}h_z^{\mu\nu}S_{xyz}^{\mu\nu}\right)}
\nonumber\\
&=&-\frac{\delta_{\mu\nu}}{2}+\frac{3}{2}\frac{\exp\left({\frac{3}{2}h_z^{\mu\nu}}
\right)\delta_{\mu\nu}}{\displaystyle\sum_{\gamma}\exp\left(\frac{3}{2}\beta
h_z^{\gamma\gamma}\right)}.
\label{msc13}
\end{eqnarray}
From this expression we see that $\langle S_{xyz}^{\mu\nu}\rangle_0=\langle S_{x+1yz}^{\mu\nu}\rangle_0=
\langle S_{xy+1z}^{\mu\nu}\rangle_0$. Since $\langle S_{xyz+a}^{\mu\nu}\rangle_0$, with $a=0,1$, or $2$, there is no dependence 
on the $x$ and $y$ variables, and we therefore we sum over these in the free energy expression as follows:

\begin{widetext}
\begin{eqnarray}
\frac{\Phi}{N^2}&=&-\frac{1}{\beta}\sum_{z=1}^N \left\{\ln
2-\frac{\beta}{2}\sum_{\mu}h_z^{\mu\mu}
+\ln\left[\sum_{\mu}\exp \left(\frac{3}{2}\beta
h_z^{\mu\mu}\right)\right]\right\}\nonumber\\
&-&J_1\sum_{z=1}^N\sum_{\mu}\left[2(M_z^{\mu\mu})^2
+M_z^{\mu\mu}M_{z+1}^{\mu\mu}\right]
-J_2\sum_z\sum_{\mu}M_z^{\mu\mu}M_{z+2}^{\mu\mu}
+\sum_z\sum_{\mu}h_z^{\mu\mu}M_z^{\mu\mu},
\label{msc14}
\end{eqnarray}
\end{widetext}
with $M_z^{\mu\nu} \equiv \langle S_{x,y,z}^{\mu\nu}\rangle_0$.

The equation defining the adjustable parameter $h_z^{\mu\nu}$ is obtained by free-energy minimization,
\begin{eqnarray}
h_z^{\mu\nu}\!\!=\!4J_1M_z^{\mu\nu}\!\! +\!J_1(\!M_{z\!+\!1}^{\mu\nu}\!+\!M_{z\!-\!1}^{\mu\nu}\!)\!
+J_2(M_{z\!+\!2}^{\mu\nu}\!+\!M_{z\!-\!2}^{\mu\nu}),
\label{msc15}
\end{eqnarray}
since $M_z^{\mu\nu}$ is a function of $h_z^{\mu\nu}$, Eq. (\ref{msc13}).
According to equation (\ref{msc13}), $M$ is traceless, consequently $\sum_{\mu}h_z^{\mu\mu}=0$,
and we use the standard parametrization,
\begin{eqnarray}
M_z=\left(
\begin{array}{ccc}
-\frac{1}{2}(Q_z+\eta_z) & 0 & 0 \\
0   & -\frac{1}{2}(Q_z-\eta_z) & 0\\
0 & 0 & Q_z
\end{array}
\right)
\label{msc16}
\end{eqnarray}
and
\begin{eqnarray}
h_z=\left(
\begin{array}{ccc}
-\frac{1}{2}(H_z+\varphi_z) & 0 & 0 \\
0   & -\frac{1}{2}(H_z-\varphi_z) & 0\\
0 & 0 & H_z
\end{array}
\right).
\label{msc17}
\end{eqnarray}
We then obtain the self-consistent equations for the order parameters;
\begin{eqnarray}
Q_z=\frac{1-\exp\left(-\frac{9}{4}\beta
H_z\right)\cosh\left(\frac{3}{4}\beta
\varphi_z\right)}{1+2\exp\left(-\frac{9}{4}\beta
H_z\right)\cosh\left(\frac{3}{4}\beta \varphi_z\right)}
\label{msc18}
\end{eqnarray}
and
\begin{eqnarray}
\eta_z=\frac{3~\text{e}^{-\frac{9}{4}\beta
H_z}\sinh\left(\frac{3}{4}\beta\varphi_z\right)}
{1+2\exp\left(-\frac{9}{4}\beta H_z\right)
\cosh\left(\frac{3}{4}\beta \varphi_z\right)},
\label{msc19}
\end{eqnarray}
with
\begin{eqnarray}
H_z\!=\!4J_1Q_z\!+\!J_1(Q_{z\!+\!1}\!+\!Q_{z\!-\!1})\!+\!J_2(Q_{z\!+\!2}\!+\!Q_{z\!-\!2})
\label{msc20}
\end{eqnarray}
and
\begin{eqnarray}
\varphi_z\!=\!4J_1\eta_z+J_1(\eta_{z\!+\!1}\!+\!\eta_{z\!-\!1})\!+\!J_2(\eta_{z\!+\!2}+\eta_{z\!-\!2}).
\label{msc21}
\end{eqnarray}

To study competing interactions we consider $J_1 > 0$ and $J_2 < 0$, 
where we defined the coupling ratio regulating the competing interactions (competition parameter), 
$p =- J_2/J_1$, whence $p$ is positive. So, from the equations (\ref{msc18}) and (\ref{msc19}), the 
free energy, $\Phi$, is given in terms of the parameters $Q_z$ and $\eta_z$ by
\begin{widetext}
\begin{eqnarray}
\frac{\Phi}{N^2J_1}
&=&-t
\sum_{z=1}^N\left[\ln2+\ln\left(\text{e}^{-\frac{3}{4tJ_1}(H_z+\varphi_z)}
+\text{e}^{-\frac{3}{4tJ_1}(H_z-\varphi_z)}
+\text{e}^{\frac{3}{2t}\frac{H_z}{J_1}}
\right)\right]\nonumber\\
&+&\frac{3}{2}\sum_{z=1}^N\left(2Q_z^2+Q_zQ_{z-1}-p~Q_zQ_{z-2}\right)
+\frac{1}{2}\sum_{z=1}^N\left(2\eta_z^2+\eta_z\eta_{z-1}-p~\eta_z\eta_{z-2}\right),
\label{msc23}
\end{eqnarray}
\end{widetext}
with $t\equiv 1/(\beta J_1)$.

\begin{figure}
\includegraphics[width=7.5cm]{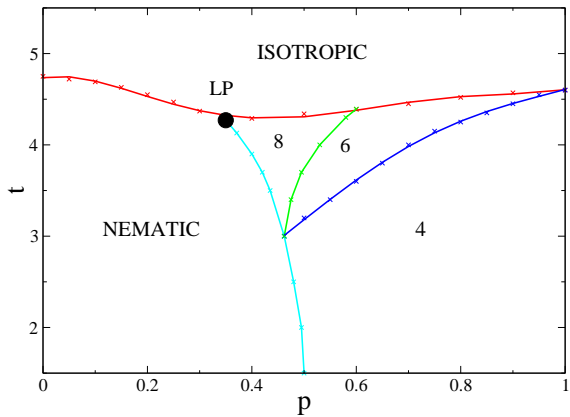}
\caption{Phase diagram of the Maier-Saupe model with competing interactions, with $J_1 > 0$, $J_2 < 0$ and 
$p = -J_2/J_1$, obtained by numerically minimizing the free energy with different imposed periodicities of the
order parameter. We have identified isotropic, nematic, and modulated (with 4, 6, and 8 periodicity) phases. 
The nematic and modulated phases meet the isotropic phase at the Lifshitz point located at 
$p \approx 0.35$ and  $t \approx 4.27$.}
\label{fig1}
\end{figure}

From Eqs.~(\ref{msc18}), (\ref{msc19}) and (\ref{msc23}) it is possible to obtain the thermodynamic phases as a function 
of the temperature and the competition parameter. To do so, at each point (t, p), we take an initial guess about the periodicity as our initial condition (here the periodicity is defined as the number of layers after which the system repeats itself. Take a simple case as an example, if molecules have only 2-states, + or -, and along z-axis, if molecules are aligned with the pattern $++--++--$..., then we say that the periodicity of this system is 4). However, eventually the initial condition converges to a final configuration based on the interitive equations, Eqs.~(\ref{msc18})-(\ref{msc21}), irrespective of the initial guess. For some cases, the final configuration may vary depending on different initial conditions. In such cases
we compare their energies based on Eq.~(\ref{msc23}) to find the ground state.

To evaluate equations (\ref{msc18}) and (\ref{msc19}) we used the iterative method (fixed points) for lattices
sufficiently large to accommodate the periodicity of each phase studied (using periodic boundary conditions). Although the competing interactions can be expected to give rise to an infinite series of of modulated phases, as in magnetic systems  
\cite{Yokoi,Philippe,Salinas,Nascimento}, the phase diagram displayed in Fig.~\ref{fig1} has been constructed by analyzing the 
free energies only of the isotropic (disordered), nematic (ordered), and modulated phases with periodicity 4, 6 and 8. 
For our purposes this is enough, since we are not aiming at describing in detail the transitions that occur between different  
modulated phases. The results already point to the existence of a Lifshitz point. We see that, beyond the isotropic-nematic 
transition, the model exhibits a transition between the nematic and modulated as well as between the isotropic and modulated 
phases. Considering $m(z)$ as the order parameter in each layer, we note that the period-4 state is special, in that the 
structure of the modulated order in a unit cell (along the $z$ axis), $m(1),m(2),m(3),m(4)$, is such that $m(1)=m(2)$ and
$m(3)=m(4)$, while for the larger periodicities $m(z)$ shows a smooth variation. Therefore, the period-4 phase 
should be considered a different ``bilayer'' phase separate from the series of modulated phases. We will confirm 
this picture with MC simulations.


\section{Monte Carlo Simulations}
\label{sec:mc}

\noindent
\begin{figure}
\includegraphics[width=7.25cm]{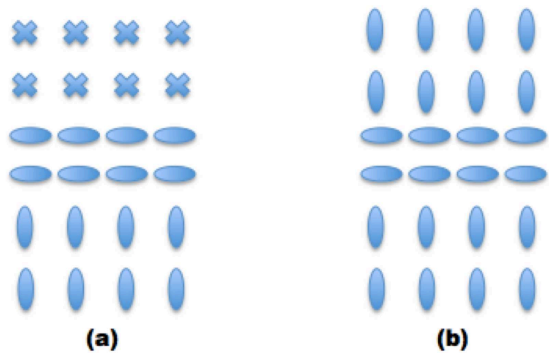}
\caption{Bilayer structure of liquid crystal molecules. (a) and (b) are two examples of the bilayer structure: (a) shows a stack of three bilayers with three different molecule orientations; (b) shows an alternating structure of bilayers with two different orientations. }
\label{bilayer}
\end{figure}

For the purpose of MC simulations we now write the effective Hamiltonian in Eq. (\ref{msc4}) as
\begin{eqnarray}
\!\!H\!=\!-J\!\sum_{(i,j)}\! ({\bf n}_{i}\cdot {\bf n}_{j})^2\!
\!+\!pJ\!\!\!\sum_{((k,k'))}\!\!\!({\bf n}_{k}\cdot {\bf n}_{k'})^2\!+\!\frac{JN(3\!-\!p)}{3}\!,
\label{newHam}
\end{eqnarray}
where ${ \bf n}_i$ is an orientational degree of freedom that can be along the directions $(0,0,1)$ or $(0,1,0)$ or $(1,0,0)$, according to the three
possible orientations of the liquid crystal molecule. The first term stands for the ferromagnetic interactions between the nearest neighbors along the
$x$, $y$, and $z$ directions, while the second term represents the interactions between the second-nearest neighbors only along the $z$-axis. 
The third term is a constant, consistent with the original Maier-Saupe model (without frustration), and $N$ is the total number of molecules; 
for a system with linear size $L$, $N=L^3$.
Compared to the Hamiltonian defined for the mean-field calculations by Eq.~(\ref{msc4}), the coupling strengths in these two Hamiltonians differ
by a factor $J=({9}/{4})J_1$, which we will adjust for later when comparing the phase diagrams. In the following, we set $J=1$ and the parameter $p$ 
will be the ratio of the two competing couplings. 

Similar to the study of 3D ANNNI model \cite{perbak}, when analyzing Eq.~(\ref{newHam}) at $T=0$ it is expected that the ground state energy
corresponds to a nematic phase when $p < 0.5$, while for $p > 0.5$, the ground state corresponds
to a modulated phase which has the bilayer structure. As seen in FIG.\ref{bilayer}, which are
example configurations of the bilayer structure. Figure \ref{bilayer}(a) shows a stack of three bilayers with three different molecule orientations,
while Fig.~\ref{bilayer}(b) shows an alternating structure of bilayers with two different orientations. There is a large ground-state degeneracy,
as any bilayer structure maintains the same lowest energy as long as two adjacent bilayers have perpendicular orientations. Therefore, at $T=0$
it is clear that $p_c=1/2$ is the transition point separating the nematic phase from the bilayer-structured phase. However, for $T>0$, it is not clear
which state the system will stay in, as the entropy plays an important role at finite temperatures. In order to draw a complete
phase diagram, we applied MC simulation to this system. In this section we will discuss the simulation method as well as the main numerical results
obtained.

In our MC simulations, we primarily consider an $8\times 8\times 8$ cubic lattice. We used a rather small size here, as this model
is very difficult to equilibrate and the simulation rapidly become much harder for larger sizes (as is well
known for frustrated systems). Nevertheless, for the purpose of obtaining a semi-quantitative view of the phase transitions of the system, we
will argue that the system size is sufficient. We have also done some calculations with a $12\times 12\times 12$ lattice and will discuss
the finite-size effects based on comparing the two sizes.
For each value of $p$, the ratio of the two competing interactions, we simulated the system at different temperatures within the range $T \in [2.5,0.5]$, 
in steps of $\Delta T=0.01$. We studied coupling ratio $p \in [0,1]$ in steps of $0.1$. At each T, we start from random initial configurations. In order to equilibrate the system, $10^{6}$ MC sweeps of $N$ 
random local updates were performed according to the Metropolis algorithm. The final results came from a bin-average of 20 bins with each bin containing the average of $10^{5}$ measurements. 

\begin{figure}
\includegraphics[width=7.5cm]{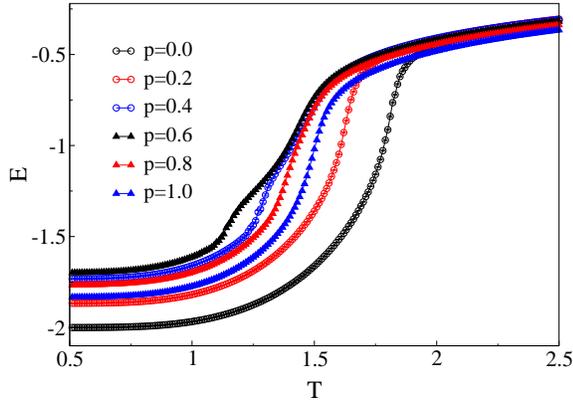}
\caption{Temperature dependence of energy density a various coupling ratios $p$ for system size $L=8$. As expected, at low temperatures, 
for system with $p<0.5$ energy approaches to $E=(2p-6)/3$, while for $p>0.5$ it approaches $E=-(9+2p)/6$. The apparent anomalies seen for
$p=0.4$ at around $T=1.4$ and $p=0.6$ around $T=1.1$ comes from the second phase transition taking place for this range of $p$.}
\label{ET}
\end{figure}

Figure \ref{ET} shows the behavior of the energy versus the temperature. As discussed previously, at low temperatures the system stays in the 
nematic phase for $p<0.5$ and the energy density (per site) then approaches to $E=(2p-6)/{3}$ when $T \to 0$, while for $p>0.5$ the system 
is in a modulated phase with bilayer structure, where the energy density approaches $E=-(9+2p)/{6}$. This behavior confirms that $p=0.5$ 
is the transition point at $T=0$.

The simplest order parameter that describes the isotropic-nematic transition is given by
\begin{eqnarray}
m=\left<\frac{1}{N}\sum_{i} \frac{(3\cos^2\theta_i-1)}{2}\right>,
\label{24}
\end{eqnarray}
where $\theta_{i}$ is the angle between the central axis of the $i$th molecule and the global director ${\bf n}$. Because of symmetry, we can simply
choose a reference director to be along $z$-direction. This order parameter $m$ easily differentiates between the nematic ($m\neq 0)$ and the
isotropic ($m=0$) phases. The transition between these two phases is known to be first-order in the standard ($p=0$) Maier-Saupe model.
The order parameter defined in Eq.~(\ref{24}) is not good for describing a modulated phase in MC simulations, however, because in modulated phases
with different orientation of the directors in different planes, it can acquire any value between $0 \le m \le 1$, depending on the values of $p$ and 
$t$, and it cannot differentiate the modulated phase from neither the isotropic phase nor the nematic phase.

To circumvent this problem, we define a layer order parameter, $m_z$, with $z \in\left[0, L-1\right]$, in each $xy$-plane along the $z$-direction;
\begin{eqnarray}
m_{z}=\left<\frac{1}{L^2}\sum_{i}^{L^2} \frac{(3\cos^2\theta_i-1)}{2}\right>_{xy-\text{plane}}.
\label{25}
\end{eqnarray}
 If there is no preferential ordering within the layers, then for finite but reasonably large system size $m_z$ is close to zero in all layers, 
 thus $m\approx 0$ and $m_z \approx 0$ defines an isotropic phase. If  $m \approx 1$ and $m_z\approx 1$, and furthermore $m_z$ has the same value 
 for all $z$, it signals a nematic phase. However, if $0<m,m_z\le1$, and at the same time $m_z$ varies in different layers, then we identify the 
 behavior as that of a modulated phase. Within the class of modulated phases, we here find two kinds: one is the bilayer-structured phase
 already discussed and illustrated with the configurations shown in Fig.~\ref{bilayer}; in this phase $m<1$ and $m_z\approx1$. The other kind
 of modulated phase has no bilayer structure and we refer to it as a single-layer modulated phase. In this case the molecules 
 align more chaotically and $m_z$ varies from layer to layer with  $0<m,m_z <1$. We associate this disorder in the $z$ direction with incommensurate 
 ordering that cannot be realized on the small lattices considered here. In addition, in this regime the behavior is clearly impacted by the
 discreteness of the director in our model. 

\begin{figure}
\includegraphics[width=7.25cm]{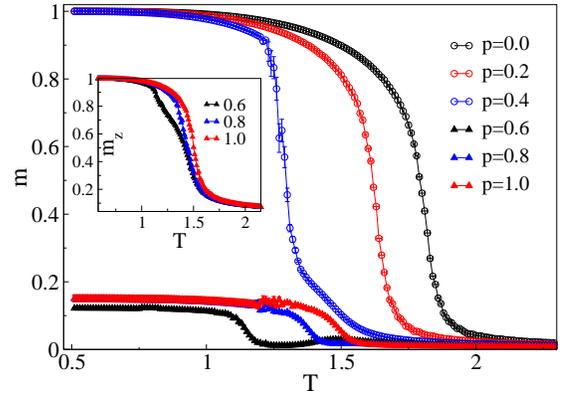}
\caption{Temperature dependence of order parameter at various ratios $p$ for system size $L=8$. The inset shows the layer order-parameter averaged
over the layers for $p \ge 0.5$.  }
\label{MT}
\end{figure}

Figure \ref{MT} shows the behavior of order parameter at various $p$ values for system size $L=8$. From high to low temperatures, systems with $p<0.5$
transition from the high-$T$ isotropic phase to the low-$T$ nematic phase, while for $p\ge 0.5$ the systems change from the high-$T$ isotropic phase to
the low-$T$ bilayer-structured phase. The inset shows the layer order parameter for three cases with $p>0.5$. The order parameter $0<m<1$ while
$m_z\approx 1$  clearly reveals the fact that in each layer the molecules align, but overall the layers align randomly along $z$-axis. However, in 
the course of the evolution from the isotropic to the nematic phase for $p<0.5$, as well as that from isotropic to bilayer structure for $p>0.5$,
we observe that, for a certain range of $p$, there is another phase that the system has to go through, which is the single-layer-structured 
modulated phase.  For if examined carefully, there are obvious abnormal behaviors of the order parameter for $p=0.4$ and $p=0.6$, in addition, we 
can also see features
beyond the statistical noise in the energy behavior at around $T=1.3$ and $T=1.1$, respectively, for the two $p$ values.  We believe that these
anomalies arise from a second phase transition in the system. For $p=0.4$, the system first undergoes the transition from the isotropic phase to
the single-layer modulated phase, and at lower temperature it goes through the second phase transition, which is from the single-layer-structured
phase to the nematic phase. For $p=0.6$, the system goes through the first isotropic to single-layer transition, followed by the second transition
from the single-layer-structured to the bilayer-structured phase. As mentioned above, this single-layer-structured phase has $m_z$ varying along
each different layer (i.e., the director is aligned differently in adjacent layers), however, it does not exhibit any aligned structure in the 
$z$-direction, in analogy with the configurations shown in Fig.~\ref{bilayer} in the case of the bilayer phase. In the following, we will provide 
more evidence for the two phase transitions and construct the phase diagram.

\begin{figure}
\includegraphics[width=7.25cm]{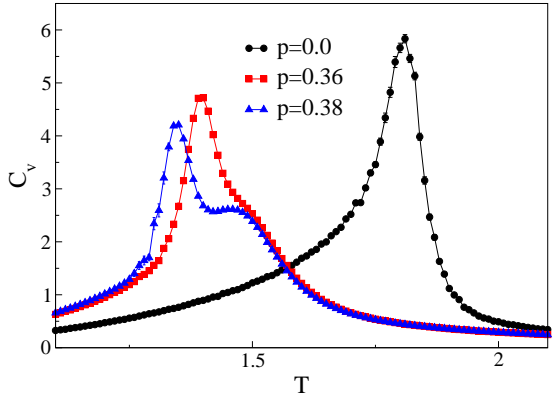}
\caption{Temperature dependence of the specific heat for $p=0.0$, $0.36$, and $0.38$. There is only one peak for $p=0$, indicating the isotropic-nematic 
phase transition, however, a second peak emerges for $p=0.36$ and becomes obvious for $p=0.38$, indicating that dual phase transitions occur at these
$p$ values}
\label{cv1}
\end{figure}

\begin{figure}
\includegraphics[width=7.25cm]{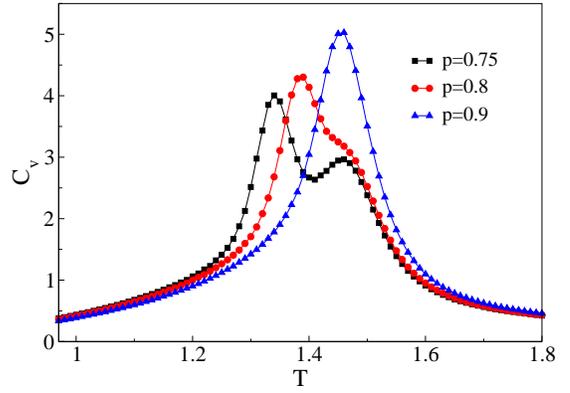}
\caption{Temperature dependence of the specific heat for $p=0.75$, $0.8$ and $0.9$. There are two peaks for $p=0.75$ and $p=0.8$, while only one peak for $p=0.9$, indicating that the dual phase transition disappears between $p=0.8$ and $p=0.9$.}
\label{cv2}
\end{figure}

 \ref{cv1} shows the behavior of specific heat $C_v$ versus temperature for several coupling ratios $p<0.5$ ($p=0.0,  0.36, 0.38$). 
Specific heat is defined as:
\begin{eqnarray}
C_v=N\beta^2(<E^2>-<E>^2), ~~~\beta=\frac{1}{T}
\end{eqnarray}
For $0.0<p<0.36$ 
we observe only a single peak, indicating that the system changes from the isotropic phase directly to the nematic phase. However, a second peak 
starts to show up for $p=0.36$ and becomes obvious for $p=0.38$, indicating that dual phase transitions appear for $p\ge0.36$, where the systems go 
through the isotropic---single-layer transition, followed by the single-layer---nematic phase transition. The crossover between the single to dual 
transitions takes place close to $p=0.36$ (within $\pm 0.01$ from this point). 
Similarly, Fig.~\ref{cv2} shows the behavior of the specific heat for several ratios $p>0.5$ ($p=0.75, 0.8, 0.9$). For $p\ge 0.9$, there is
only one peak, indicating that the systems change from the isotropic phase directly into the bilayer-structured phase. However, for $0.5<p\le0.8$ 
the specific heat again exhibits two peaks, indicating that systems go through the dual isotropic---single-layer---bilayer phase transitions.  
The crossover here occurs at  $p\approx0.8$.

\begin{figure}
\includegraphics[width=7.25cm]{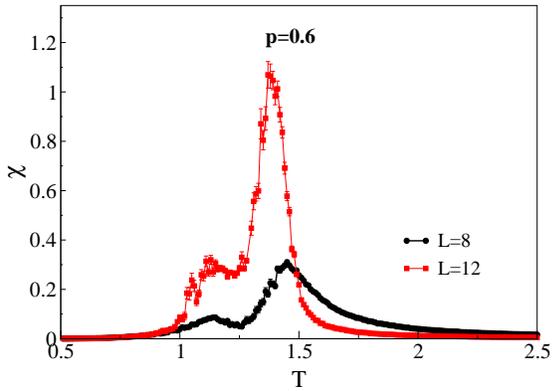}
\caption{Temperature dependence of $\chi$ at $p=0.6$ for L=8 and L=12. The peak near $T\approx1.44$ indicates the isotropic---single-layer phase transition, while
  the anomaly at around $T=1.14$ corresponds to the transition from single-layer-structure to bilayer-structure. Comparing the results for the
two cases, we can see the finite-size-effect playing some role in the $L=8$ systems, but we believe
that the overall effects on the phase boundaries are minor.  }
\label{chi}
\end{figure}

Not only does the specific heat shows evidence for dual phase transitions, but there are also corresponding anomalies in the behavior of the
order-parameter fluctuations. In analogy with the susceptibility in a magnetic system, we define a `susceptibility' $\chi$ for our model as
\begin{eqnarray}
\chi = N\beta \left(\langle m^2\rangle-\langle m\rangle ^2\right),
\label{26}
\end{eqnarray}
and this quantity should diverge at any of the ordering transitions discussed. Fig.\ref{chi} shows the behavior of the susceptibility versus 
temperature at $p=0.6$ for both $L=8$ and $L=12$. In both cases, two well-separated peaks can be seen, indicating the two phase transitions take 
place with both lattices. Moreover, comparing the results for the two cases, we can see that the peaks become higher and narrower with increasing 
size, as expected for peaks diverging in the thermodynamic limit, and the peak positions shift by only about $5\%$. While the finite-size-effect is, 
thus, playing some role in the $L=8$ systems, we believe that the overall effects on the phase boundaries are minor. 

\begin{figure}
\includegraphics[width=7.5cm]{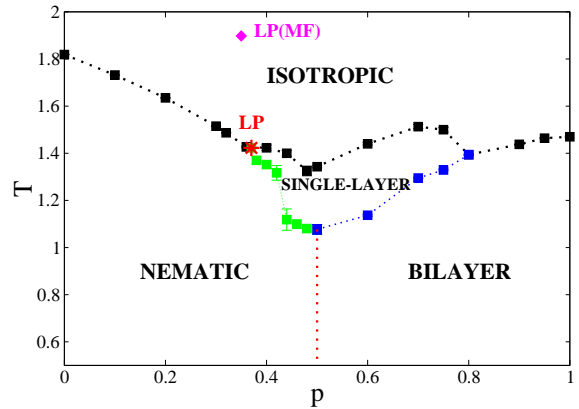}
\caption{Phase diagram of the discrete Maier-Saupe model with competing interactions, obtained on the basis of MC simulations with $8\times8\times8$ 
lattices. The red star point indicates the Lifshitz point (LP) based on our MC results, while the magenta diamond shows the LP based on Mean-Field 
results in Fig.~\ref{fig1} adjusted by the factor $9/4$ relating the two versions of the Hamiltonian.}
\label{mc_diagram}
\end{figure}

We finally, in Fig. \ref{mc_diagram}, present the phase diagram drawn based on our MC simulations. By fitting the specific heat results to high-order 
polynomials in the peak regions, we can locate the transition temperature $T_c$ for various values of $p$. We estimate error bars by fitting multiple 
times through the bootstrapping method. As discussed above, the system has four different phases: isotropic phase, nematic phase, single-layer-structured 
modulated phase as well as bilayer-structured modulated phase. For a certain range of $p$ ($0.36<p<0.8$), the system goes through two phase 
transitions as the temperature is lowered, as it has to go through the single-layer-structured modulated phase as an intermediate state before 
reaching the nematic state or the bilayer-structured state at low temperature. The red star in the phase diagram represents the Lifshitz point (LP), 
located at $p\approx 0.36$, $T\approx 1.42$. Recall that the Mean-field calculations found LP at $p_{MF} \approx 0.35$ and  $t_{MF} \approx 4.27$, 
since the coupling strengths in the two Hamiltonians differs by a factor of ${9}/{4}$, converting $t$ used there to the the temperature defined in 
the MC simulation, we obtain $T_{MF}=(4/9)t_{MF} \approx 1.90$. The magenta diamond in the MC phase diagram shows the LP based on mean-field 
calculation for comparison.

According to the phase diagram of the 3D ANNNI model from MC simulation, it is known that the paramagnetic-ferromagnetic phase transition is continuous, while both the para-modulated and ferro-modulated transitions are first-order \cite{murtazaev}. For our model, which is equivalent to 3D 3-state ANNN-Potts model, the isotropic-nematic phase transition is first-order (just as the case in the standard 3D 3-state Potts model). For the nemetic-modulated and isotropic-modulated phase transitions, we are not certain but we think they are most likely first-order phase transitions as well. Potentially if we have data from more system sizes, we can test this conclusion through finite-size analysis.

Compared to the phase diagram from the mean-filed calculations (Fig.\ref{fig1}), the MC phase diagram shows no clear signs of modulation in the 
modulated phase (in the case of the single-layer structure as well as the double-layer structure). Instead, the inter-layer orientation always appears 
random and we cannot detect any meaningful correlations. Most likely, this is an indication of incommensurate ordering pitch that cannot be realized on 
the small lattices considered here and with the discreteness of the director. Nevertheless, the regime marked as phase with periodicity 4 in the 
mean-field phase diagram represents exactly the bilayer structure in Fig. \ref{bilayer}(b). In this sense, despite some quantitative difference, in 
general the two phase diagrams are very consistent with each other.

\section{Final Considerations}
\label{sec:summary}

\noindent

In this work we have investigated the phase diagram of a 
extended version of the discrete Maier-Saupe model with competing interactions
between nearest and second-nearest neighbors in one direction. The model also corresponds to 
a $3$-state Potts version of the ANNNI model.
 Initially we carried out the studies 
by means of mean-field calculations. Even with a variational mean-field approach, the competing interactions produce a phase diagram with modulated 
structures. By applying numerical methods to find the order parameter, we obtained the transition lines between isotropic-nematic, isotropic-modulated, 
and nematic-modulated phases. To compare with the mean-field results we also employed MC simulations. In this case, the order parameter in 
Eq.~(\ref{24}) is not able to distinguish the modulated phase of the isotropic and nematic phases, thus we introduced a modified layer-order parameter 
to distinguish the phases in more detail. With the MC results, we were then able to identify four phases of the system and construct the full phase 
diagram. In addition to a nematic-isotropic transition, which are present in the absence of competing interactions, the model shows both transitions 
between isotropic-modulated and nematic-modulated phases. Although one can not expect the mean-field phase diagram to be quantitatively correct, 
the MC simulations still corroborate the general pattern of the mean-field phase diagram, and even quantitatively the Lifshits point 
appears almost at the same coupling ratio as in the mean-field phase diagram, and at a temperature only about $25\%$ lower.

We stress that, even with the discretized version of the Maier-Saupe model considered here, its phase diagram shows an interesting rich structure. 
In order to make definite statements about the relevance of our results to LCs, the model should be extended to continuous degrees of freedom;
classical Heisenberg spins taking continuous values over a unit sphere. We regard our study of the discrete model as a first step on the path to future 
studies of frustrated models of LCs. Most likely the modulated phases we have found here will survive with continuous degrees of freedom, though 
details such as the phase boundaries and the pitch of the modulation with the frustration parameter and the temperature may shift. We also expect that 
such a more refined model might be able to capture elements of more complex liquid crystalline phase transitions, such as the isotropic-smectic 
C* and isotropic-cholesteric transitions. In this last case, it should be pointed out that it is necessary to further generalize the Hamiltonian Eq. (\ref{msc4})  
including odd-chirality terms, which are important to take into account the striking feature of handedness observed in the cholesteric phase.

\section{Acnowledgments}

\noindent
We would like to thank P. Gomes,  R. Kaul,  G. Landi,  M. Oliveira, R. Oliveira, and S. Salinas for useful discussions and suggestions. P.F.B was supported
by Funda\c c\~ao de Amparo a Pesquisa do Estado de S\~ao Paulo (FAPESP) and the Condensed Matter Theory Visitors Program at Boston University.
N.X. and A.W.S. were funded in part by the NSF under grant No.~DMR-1410126. Some of the calculations were carried
out on Boston University's Shared Computing Cluster.


\medskip

\end{document}